\documentstyle[12pt]{article}
\begin{document}

\thispagestyle{empty}
\newcommand{\beq}{\begin{equation}}
\newcommand{\eeq}{\end{equation}}
\newcommand{\RA}{\rightarrow}
\newcommand{\lA}{\vert l \vert}

\baselineskip=0.6cm
\begin{flushright}{.}\end{flushright}
\begin{flushright}{TIT/HEP-273}\\
{November 1994}\end{flushright}
\begin{center}
\vspace{15mm}

{\large\bf The  Casimir energy for two-dimensional deformed sphere}\\

\bigskip
\vspace{0.3in}

{N.~Shtykov}\\
\medskip{\it Department of Physics , Tokyo Institute of
Technology},\\
{\it Oh-Okayama, Meguro-ku, Tokyo, 152, Japan}$^{\dagger}$\\
\medskip{\it and}\\
\medskip{\it D.~V.~Vassilevich}\\
\medskip{\it Department of Theoretical Physics, St. Petersburg
University,}\\
{\it 198904 St.Petersburg, Russia}\\

\end{center}
\vspace{15mm}

\centerline{\bf Abstract}
\begin{quotation}
We compute the Casimir energy for a free scalar field on the
spaces $\,R^{m+1}\,\times\,\tilde S^2\,$ where $,\tilde S^2\,$
is two-dimensional deformed two-sphere.

\end{quotation}

\vfill
\noindent
$^{\dagger}$ On leave from: {\em Irkutsk university, Russia};\\
Electronic mail: shtykov@phys.titech.ac.jp

\newpage

The explanation of the Casimir effect lies in the fact, that quantum fields respond to the presence of external constraints. Meaningful constraints can be
defined, for example, in terms of the boundary conditions which fix
these fields.
In a number of works, the Casimir energy has been calculated in $d$-dimensional
spaces  for free scalar,
spinor and vector fields at a zero and non-zero temperature with the
Dirichlet, von Neumann and Robin boundary
conditions and the periodicity condition (compactification $R^1$ to $S^1$)
\cite{1}-\cite{6}. ( For review of several
applications of the Casimir energy see ref. \cite{7}).

The calculation of the one-loop effective potential ( which is closely
related to the Casimir energy) in product spaces $R^{n}\times Y$ where
$Y$ is a compact manifold was motivated by the Kaluza-Klein models and involves
some mathematical problems related to regularization of the functional
determinants , $\det \Delta$ with the eigenvalues of a second-order
differential operator $\Delta$ \cite{8}. The scalar functional determinants
have been also  determined in terms of $\zeta$ functions on
sectors of the two-dimensional disk and spherical cap \cite{9}.
Recently, there has been some  interest in
investigating spaces with conical singularities. Physical applications for
such spaces range
from quantum cosmology \cite{10} and cosmic strings \cite{11} to finite
temperature field theory \cite{12}. Some mathematical aspects were
examined in the papers \cite{13}. Typically, two alternative cases are
considered. In the first case, the conical singularities come from
factorization of a sphere $S^d$ with respect to a finite group
$\Gamma$, which acts on $S^d$ with fixed points. On this way,
only discrete series of conical singularities can be studied.
In the second case, a standard two-dimensional cone with
varying angle deficit is introduced. However, in this approach
curvature can be considered only as perturbation of the flat case.
We propose to consider a two-dimensional sphere with two conical
singularities with  an arbitrary angle parameter. This can be treated
as an imitation of the effects \cite{12} for essentially curved space
without exact relevance to any particular physics. In this sence
our aim is two-fold. First, we study technical aspects of the
problem and second, we explore qualitative behaviour of the effective
potential which appears to be quite different from the non-singular
case.

In this letter we compute the vacuum energy for free scalar
fields with a background geometry of $\,R^{m+1}\,\times\,\tilde
S^2\,$ where  $\,R^{m+1},$ is  the $\,m+1$-dimensional
Euclidean space, $\,\tilde S^2\,$ is the two-dimensional compact
manifold constructed from two-sphere $\,S^2\,$ by the deformation of
its metric. We find the exact eigenvalues of the Laplace operator and
compute the vacuum  energy as a function of the deformation parameter by
making use of the zeta-function technique. In odd-dimensional spaces
$\,R^{1}\,\times\,\tilde S^2\,$ and $\,R^{3}\, \times\,\tilde S^2\,$,
we obtain the finite vacuum energy, which changes its sign with some value of
the deformation parameter. In even dimensional spaces $\,R^{2}\,\times\,\tilde
 S^2\,$ and $\,R^{4}\, \times\,\tilde S^2\,$ the vacuum energy
contains a divergent part and has the following form
\beq
E\,=\,\frac{1}{\epsilon}f_{0}(m,\rho)\, +\,f_{1}(m,\rho)\,+\,f_{0}\log(r\mu)
\eeq
where $r$ is the radius of $\tilde S^2\,$ and $\mu$ is an arbitrary scale
parameter introduced with zeta regularization. We compute the functions
$f_{0}$ and $f_{1}$ with some range of the deformation parameter $\rho$.
It is interesting to note that in $\,R^{4}\, \times\,\tilde S^2\,$ the
divergent part in (1) dissappears with $\rho\,\simeq 0.415$ yielding the
finite value
$\,E\,\simeq 1.47\ 10^{-5}$. (Here and throughout the paper we set
$\,r\,=1\,$). We compare our results with those obtained in flat
spaces for a scalar field satisfying in two dimensions to the Dirichlet, von
Neumann and periodic boundary conditions.

Let us begin with the scalar Laplacian eigenvalues on a deformed sphere
$\tilde S^2$ . The metric on $\tilde S^2$ can be expressed in the form
$$
ds^2=dx_0^2+\,sin^2x_0\,\rho^{-2}d\psi^2 ,\quad 0 \le \psi
\le 2\pi$$
where $\rho$ is a positive constant parameter. This space can be also
obtained by identification of points with polar angles $\alpha$
and $\alpha +2\pi /\rho$ on round $S^2$ with a maximally symmetric
metric. The $\tilde S^2$ has two singular points $x_0 =0$ and $x_0=\pi$.
Let us perform the Fourier decomposition with respect
to the coordinate $\psi$.
\beq
\phi (x_0,x_i)=\sum_{l=-\infty}^\infty f_{(l)}(x_0)
\exp (il\psi) \eeq
Substituting the decomposition (2) in the eigenvalue
equation
$$ \Delta \phi =-\lambda \phi $$
we obtain the following ordinary differential
equation
\beq [\partial_0^2 + {\rm ctg }x_0 \partial_0
-\frac {\rho^2 l^2}{\sin^2 x_0} ]f_{(l)}=-\lambda f_{(l)}.
\eeq
Equations of this type were considered in our
previous papers \cite{14}.
We shall drop the subscripts $(l)$ for a while.
Let us change the independent and dependent variables
$$ f=h \sin^{\lA \rho} (x_0), \quad
z=\frac 12 (\cos x_0 +1). $$
The $\lA$ denotes absolute value of $l$.
The equation (3) then takes the form
\beq z(z-1)h''+2(\rho \lA +1)(z-\frac 12 )h'+(\rho \lA (
\rho \lA +1) -\lambda )h=0 . \eeq
Prime denotes  differentiation with respect to $z$.
According to the general prescription \cite{16} let us
express $h$ as the power series
\beq h(z)=\sum_{k=0} \alpha_k z^k.
\eeq
Substitution of (5) in (4) gives a recurrent condition
on the coefficients $\alpha_k$
\beq \alpha_{k+1}=\alpha_k
\frac {k(k-1)+2(\rho \lA +1)k+\rho \lA (\rho \lA +1) -\lambda}{
(k+1)(k+\rho \lA +1)}. \eeq
The denominator of (6) is positive for all $k$. The
eigenfunctions $h_k$ can be found by imposing the
condition on the numerator of (6) to be equal to
zero. We obtain  the eigenvalues
\beq \lambda_{l,k}=(k+\rho \lA +\frac 12 )^2 -\frac 14  \eeq
where we restored the dependence on the index $l$. ( Eq. 7 has been also
considerd in Ref. \cite{15}).

In the case of the unit round sphere $ S^2$ with
$\rho =1$ we obtain from (7)
$$ \lambda_{(l)k}=-(k+\lA )(k+\lA +d)=-n(n+d), \quad n=k+\lA
 $$
Thus  equation (6) reproduces the correct eigenvalues
of the scalar Laplace operator on the unit round
$S^2$. One can also verify that the degeneracies
have  the correct values.

With zeta regularization the Cazimir energy density for a scalar field on
$\,R^{m+1}\times \tilde S^2\,$ can be written as
\beq
E  = \lim_{s\RA -1}\;\frac{\Gamma{((s-m)/2)}\,\mu^{2\epsilon}}{2(4\pi)^{m/2}\,
\Gamma(s/2)}\,\Bigl (2\sum_{l=1}^{\infty}\,\sum_{k=0}^{\infty}\,\lambda_{l,k}^
{-s/2}\,+\,\sum_{k=0}^{\infty}\,\lambda_{0,k}^{-s/2}\,\Bigr )
\eeq
where $\epsilon\;=\;(1+s)/2$.
Making use of the Hermite formula \cite{16}
$$
\zeta(z,q) = \frac{q^{-z}}{2} + \frac{q^{1-z}}{z-1} + 2\int_{0}^{\infty}
dx\,\sin(z\,\arctan(x/q))\frac{(q^2 + x^2)^{-z/2}}{e^{2\pi x} - 1}
$$
and taking the expansion
$$
\sum_{n=0}^\infty (n+q)^l((n+q)^2+M)^{\frac{m+s}{2}}\,=\,\sum_
{ k=0}^\infty\;\frac{(-1)^k\Gamma(k-(m+s)/2)}{\Gamma(-(m+s)/2)k!}\,
\zeta(2k-l-m-s,\,q)\,M^k
 $$
we rewrite (8) as
\beq
E  = \lim_{s\RA -1}\;\sum_{r=0}^{\infty}\,\frac{\Gamma{(r+(s-m)/2)}\,\mu^{2\epsilon}}{2(4\pi)^{m/2}\Gamma(s/2)4^r r!}\,\Bigl ( P(s-m+2r,\rho)\,+\,F(s-m+2r,\rho)\,\Bigr )
\eeq
where
 $$
P(z,\rho)\,=\,\rho^{-z}\zeta(z,1+1/(2\rho))\, +\,\frac{2}{z-1}\rho^{1-z}\zeta(z-1,1+1/(2\rho))\, +\,\zeta(z,3/2),
 $$
 $$
F(z,\rho)\,= \,2\sum_{p=0}^{\infty}\,(-1)^{p+1}c_p(z)\,\sum_{n=0}^{\infty}
\,\frac{\Gamma (n+z/2)}{\Gamma (z/2)n!}\rho^{-2p-2n-z-1}\, $$
 \beq
\times \zeta(2p+2n+z+1,1+1/(2\rho))\zeta(-2p-2n-1).
\eeq
The coefficients $c_p$ are determined from
$$
\sin(z\,\arctan(x))\, =\,\sum_{p=0}^{\infty}\,c_p(z)\,x^{2p+1}
$$
The expression (9) involves the divergent terms coming from the poles of
$\Gamma(z)$ at $\;z = -k\ \ (k =0,1,2...)\;$
\beq
\Gamma{(-k +\epsilon)}\;= \frac{(-1)^k}{k!}\,(\frac{1}{\epsilon}
+ \psi(k +1) + O(\epsilon))
\eeq
and the pole of  $\,\zeta(z,q)\;$ at $\,z=1\,$
\beq
\zeta(1 + 2\epsilon, q) = \frac{1}{2\,\epsilon} - \psi(q) + O(\epsilon).
\eeq
However, in odd dimensions the divergent term $\;1/(z-1)\;$ at $\,z=1\,$
cancels the pole of zeta function yielding the finite Casimir energy. We find,
omitting the details, from (9)-(12) the following numerical values  for the
Casimir energy $E_m$.
$$  \ \ \ \ \ \ \ \ \rho\,=\,0.4\ \ \ \ \ \ \ \ \ \ \ \ \rho\,=\,0.8
 \ \ \ \ \ \ \ \ \
\ \ \ \ \ \ \ \ \rho\,=\,1\ \ \ \ \ \ \ \ \ \ \ \ \ \ \ \rho\, =\,2$$
$$ E_0 \ \ \ \ \ \ \ 0.07385\ \ \ \ \ \ \ \ \ -0.09987\ \ \ \ \ \ \
\ \ \ \ \ \ -0.13246\ \ \ \ \ \ \ \ \ -0.1990, $$
\beq
E_2 \ \ \ \ \ \ \ 16.1\ 10^{-4} \ \ \ \ \ \ \ \ -1.53\ 10^{-4}\ \ \ \ \ \ \
\ \ \ \ \ \ -4.78\  10^{-4}\ \ \ \ \ \ \ \ \ -11.4\ 10^{-4}.
\eeq

In even dimensions,  for $E_m$  given in the form (1) we compute both $f_0(m,
\rho)$ and $f_1(m,\rho)$.\\
For $m=1$
$$  \ \ \ \ \ \ \ \ \rho\,=\,0.4\ \ \ \ \ \ \ \ \ \ \ \ \rho\,=\,0.8
 \ \ \ \ \ \ \ \ \
\ \ \ \ \ \ \ \ \rho\,=\,1\ \ \ \ \ \ \ \ \ \ \ \ \ \ \ \rho\, =\,2$$
$$ f_0(1,\rho) \ \ \ \ \ \ \ -0.00144\ \ \ \ \ \ \ \ \ -0.00216\ \ \ \ \ \ \
\ \ \ \ \ \ -0.00265\ \ \ \ \ \ \ \ \ -0.0063, $$
\beq
f_1(1,\rho) \ \ \ \ \ \ \ -0.0023 \ \ \ \ \ \ \ \ -0.0097\ \ \ \ \ \ \
\ \ \ \ \ \ -0.0117\ \ \ \ \ \ \ \ \ -0.0179.
\eeq
For $m=3$
$$ f_0(3,\rho) \ \ \ \ \ \ \ 1.2\ 10^{-6}\ \ \ \ \ \ \ \ \ -2.56\ 10^{-5}
\ \ \ \ \ \ \ \ \ \ \ \ \ -4.02\ 10^{-5}\ \ \ \ \ \ \ \ \ -21.8\ 10^{-5}, $$
\beq
f_1(3,\rho) \ \ \ \ \ \ \ 2.81\ 10^{-5}\ \ \ \ \ \ \ \ \ -6.18\ 10^{-5}
\ \ \ \ \ \ \ \ \ \ \ \ \ -9.15\ 10^{-5}\ \ \ \ \ \ \ \ \ -18.1\ 10^{-5}.
\eeq

 Now some notes are in order. To extract the effect of deformation
we should redefine $r$ such that the volume on $\tilde S^2$ is the same as that
of $S^2$ with a unit radius. In this case, we have to consider the quantity
$\bar E_m\,=\,E/\rho^{m+1}$. It is easy to see from (13) that $\bar E_0\,$ and
$\bar E_2\,$ take the minimal values at $\;\rho\,=\,1$. Thus, sphere $S^2$
tends to keep its shape at least for the slight deformation. A detailed study
 shows that this is the only minimum of $\bar E_m\,$. The asymptotic behaviours
of $\bar E_m\,$ ($m=0,2$)are
$$
\bar E_m\,\sim\,\rho^{-m-2},\ \ \ \ \ \ \rho\RA 0 ,$$
 $$
\bar E_m\,\sim\,-\rho^{-m-1}\;\log\rho\ \ \ \ \ \
\rho\RA\;\infty .
 $$
Indeed, the singularity at $\rho\;=\;0$ gives rise to the repulsive
force, tending to expand $\tilde S^2$.
Since $\bar E_1\,$ and $\bar E_3$ have the divergences and depend on
an arbitrary scale $\mu$ , some renomalization procedure should be used to
obtain the finite result. However, one can see from (15)  that $f_0(3,\rho)$
becomes equal to zero with the certain value of $\rho\;=\;\rho_0$ in the
region
$\,0.4\;<\rho \;<0.6\;$ and the divergent part is removed from $E_3$.
This reflects the fact that the conformal anomaly in  $R^4\times \tilde
S^2$ is canceled with the deformation parameter $\rho_0$ .

Finally, we compare the results in (13) with those obtained for a scalar
massless field in  general three- and five-dimensional hyper-cuboidal
regions  with two sides of finite length $a_1\,,a_2$ and one and three sides
with length $L\gg a_1\,,a_2$ respectively. In the case of periodic boundary
conditions the vacuum energy as a function of $a_2/a_1$ has the minimal
value when $a_2\,=\,a_1$ and is always negative. The energy
obtained with von Neumann boundary conditions \cite{2} also exhibits this
behaviour. With Dirichlet boundary
conditions the energy takes the positive maximal values at $a_2/a_1\,=\,1$
for both three- and five-dimensional spaces and tends to negative infinity
when $a_2/a_1\,\gg 1$ \cite{2}. Thus, we can see that in all these models
the behaviour of the vacuum
energy  is quite different from that of the energy in  $R^1\times \tilde
S^2$ and $R^3\times \tilde S^2$ spaces.\\

{\bf Acknowledgements}

This work was partially supported by the Russian Foundation
for Fundamental Studies, grant 93-02-14378. One of us (DV)
is grateful to Ignati Grigentch for colaboration on
Laplace operator on deformed spheres and to Sergio Zerbini,
Guido Cognola and Luciano Vanzo for hospitality at Trento
and discussions. We are also grateful to D. Fursaev for pointing
out Ref. \cite{15}.

\end{document}